\documentclass{jfm}
\usepackage{graphicx}
\usepackage{epstopdf, epsfig}
\usepackage{amsmath}
\usepackage{graphicx}
\usepackage{multirow}
\usepackage{caption}
\usepackage{subcaption}
\usepackage{float}
\usepackage[utf8]{inputenc}
\usepackage{array}
\DeclareGraphicsExtensions{.pdf,.png,.jpg,.eps}
\usepackage{epstopdf}
\usepackage{xcolor}
\usepackage{multimedia}

\shorttitle{Stability of an inviscid flow through a tube with porous wall}
\shortauthor{R. Patne}

  \title{Stability of an inviscid flow through a tube with porous wall}
   \author{Ramkarn Patne\aff{1}\corresp{\email{ramkarn@che.iith.ac.in}}}

 \affiliation{\aff{1}Department of Chemical Engineering, Indian Institute of Technology Hyderabad, Kandi, Sangareddy, Telangana 502285, India}

\begin{document}

\maketitle

\begin{abstract}
 
 The temporal stability of an inviscid flow through cylindrical geometries with a porous wall subjected to non-axisymmetric perturbations is investigated in the present work using an unsteady Darcy equation for the porous layer. An expression for the perturbation energy exchange term between the fluid and porous layers is derived by integrating the perturbation equation for the porous layer which is then used to prove the propositions. The necessary and sufficient condition for the existence of the instability in flows through cylindrical geometries  is shown to be $\frac{d}{dr} \left( \frac{r \frac{dV}{dr}}{n^2+k^2 r^2} \right) (V-V_I)<0$ somewhere in the flow where $V$ and $V_I$ are the axisymmetric base-state velocity profile and velocity at the fluid-porous layer interface, respectively. The parameters, $k$ and $n$ are the axial and azimuthal wavenumbers, respectively. Additionally, the bounds on the phase speed ($c_r$) and growth rate ($k c_i$) are derived. 
 
 The derived propositions are used to determine the stability of the sliding Couette, annular Poiseuille, and  Hagen-Poiseuille flows with a porous wall. The analysis reveals, for the first time, the existence of both axisymmetric ($n=0$) and non-axisymmetric ($n\neq 0$) modes of instability in all three flows. The Hagen--Poiseuille flow through a tube with the non-porous wall is linearly stable. Therefore, the instability predicted here is solely due to the porous nature of the wall. The mechanism responsible for the predicted instability is the exchange of the perturbation energy between the fluid and porous layers.

\end{abstract}


\section{Introduction}

 Flows past a porous layer saturated with same fluid are encountered in natural and industrial settings \citep{Childs-Collis-George-1950,Discacciati-et-al-2002,Goyeau-et-al-2003,Liu-et-al-2008,Hill-Straughan-2008,Langre-2008,Lyubimova-et-al-2016}.  \cite {Nield-1977,Nield-1983,Chen-Chen-1988,Nield-1991} studied the thermal convection of a fluid overlying a porous layer. \cite{Chen-Chen-1988} predicted that there exists a bimodal structure of the neutral stability curves depending on the depth ratio. The depth ratio in their analysis was defined as the ratio of the fluid layer thickness to the thickness of the porous layer.

 The linear stability of planar Couette and Poiseuille flows past a porous layer has been extensively studied by \cite{Sparrow-et-al-1973,Denga-Martinez-2005,Chang-et-al-2006,Tilton-Cortelezzi-2006,Hill-Straughan-2008,Tilton-Cortelezzi-2008,Chang-et-al-2017,Samanta-2017} using both the unsteady Brinkman and Darcy models.  \cite{Chang-et-al-2006} utilised the unsteady Darcy model for the porous layer to study the linear stability and predicted three modes of instability. These three modes of instability were classified based on the origin of the mode as fluid layer mode, porous layer mode, and interface mode of which the last mode was believed to be originating due to the shear layer at the interface.   \cite{Tilton-Cortelezzi-2006,Tilton-Cortelezzi-2008}  studied the effect of the porous layer permeability on the instabilities. Recently, \cite{Chang-et-al-2017}  showed the stabilizing effect of the Couette component on the instabilities originating due to the plane Poieuille flow past a porous layer. \cite{Sparrow-et-al-1973,Denga-Martinez-2005,Chang-et-al-2006,Tilton-Cortelezzi-2006,Hill-Straughan-2008,Tilton-Cortelezzi-2008,Chang-et-al-2017,Samanta-2017} employed a two-domain approach, i.e., fluid and porous layers were treated as separate layers being separated by an interface. \cite{Silin-et-al-2011,Ghosh-et-al-2019} employed the one-domain continuum approach to study the linear stability analysis of a flow overlying a porous medium. In the one-domain approach, the volume average Navier--Stokes equations are used to describe the flow in both fluid and porous layers.

  The previous studies concerning the stability of the flows past a porous layer considered planar flows. In geophysical flows \citep{Li-Chen-2012}, physiological processes \citep{Bates-Harper-2002,Reyes-et-al-2008}, and industrial applications \citep{Pangrle-1991,Nassehi-1998,levenspiel-1999,Tilton-2012} flows in cylindrical geometries with a porous wall are relevant. However, to the best of the author's knowledge, there has not been a study concerning the stability of such a class of flows. The present study fills this important void albeit in the inviscid flow limit. The present analysis is also relevant for entry flow in a cylindrical geometry with a porous wall or the flow geometries with non-uniform cross-sections, the latter being a common feature of the geophysical flows. For the entry flow, the velocity profile may not be fully developed or for flow geometries with slowly varying cross-sections, the flow could be different from the flow geometries with uniform cross-sections. The stability analysis of such flows will be considerably mathematically intricate and computationally expensive due to the unavailability of an analytical velocity profile. Thus, consideration of an arbitrary velocity profile to obtain meaningful conclusions about the stability of flows could be of interest thereby making the present analysis relevant. Furthermore, \cite{Ali-et-al-2017,Ali-et-al-2018b,Ali-et-al-2018} have demonstrated the use of a porous layer to passively or semi-actively control the flow-induced noise
and vibrations. These applications often involve flows at a high Reynolds number which may justify the use of an inviscid flow and thus the present analysis.

 The qualitative features of the stability of planar inviscid flows past rigid surfaces can be obtained from classical analyses of \cite{Fjortoft-1950,Rayleigh-1880,Hoiland-1953}. \cite{Squire-1933} proved that for flows past rigid surfaces, two-dimensional perturbations become unstable at lower Reynolds number than corresponding three-dimensional perturbations. Thus, the  equation governing the stability of inviscid parallel flows can be readily derived to result in a second-order ordinary differential equation which is also known as the `Rayleigh equation' \citep{drazinreid}. However, for flows in cylindrical geometries, the theorem of \cite{Squire-1933} is not applicable. Thus, in the present study, non-axisymmetric perturbations are assumed for the analysis.

 Using the Rayleigh equation, \cite{Fjortoft-1950} showed that the necessary and sufficient condition for the existence of the instability in planar flows is $D^2V (V-V_{inflexion})<0$ somewhere in the flow where $V$ is the base-state velocity profile, $V_{inflexion}$ is the velocity at the point of inflexion, and $D^2$ denotes the second-order derivative of $V$.  An equivalent of the \cite{Fjortoft-1950} theorem for the flows in cylindrical geometries is due to \cite{Maslowe-1974}. It must be noted that the condition derived by \cite{Maslowe-1974} is much more general since his criterion was originally derived for the rotation imposed on the Hagen--Poiseuille flow. \cite{Lalas-1975} further extended the analysis of \cite{Maslowe-1974} to compressible flows. In addition to proving the condition for the instability, he also derived the bounds on the growth rate of the perturbations. The present analysis does not consider the imposed rotation thus the predictions obtained here will be applicable to the flow geometries having base-state velocity in the axial direction. Another classical result is due to \cite{Hoiland-1953} who showed that there exists an upper bound on the growth rate of the unstable perturbations in planar flows. The bound being the maximum of the absolute value of the base-state velocity gradient. 
 
 An extension of the classical results of \cite{Fjortoft-1950,Rayleigh-1880,Hoiland-1953} for the planar flows past deformable surfaces was studied by \cite{yeo87,yeo94a}. Their analysis showed that the result of \cite{Hoiland-1953} holds true even for the flows past deformable surfaces. However,  \cite{Fjortoft-1950} criterion is modified to $D^2V (V-V_w)<0$ where $V_w$ is the fluid velocity at the wall. The results of \cite{Maslowe-1974,Lalas-1975} were extended to the Hagen--Poiseuille flow through a deformable tube by \cite{kumaran96} and \cite{shankar-kumaran-2000} to axisymmetric and non-axisymmetric perturbations, respectively. \cite{kumaran96} showed the absence of the unstable axisymmetric inviscid modes for the Hagen-Poiseuille flow through a deformable tube. However, \cite{shankar-kumaran-2000} showed that unstable inviscid modes can exist for non-axisymmetric perturbations in Hagen-Poiseuille flow through a deformable tube thereby showing the importance of considering the non-axisymmetric perturbations. Numerical analysis by \cite{shankar-kumaran-2000} further affirmed the predictions of \cite{kumaran96,shankar-kumaran-2000}. The methodology used here to prove the propositions is similar to \cite{yeo87,yeo94,kumaran96,shankar-kumaran-2000} thus, based on a proven procedure.

 A fluid flowing past a porous layer leads to a perturbation energy exchange between the flowing fluid and the flow inside the porous layer via the fluid-porous interface. This energy exchange then alters the stability behaviour of the flowing fluid. Here, an expression for such exchange of perturbation energy is derived which is then expressed in terms of the fluid quantities. The perturbation exchange term is next utilised to obtain qualitative conclusions about the stability of the whole system.

The rest of the paper is arranged as follows. The governing equations for non-axisymmetric perturbations, boundary conditions, and perturbation energy exchange term responsible for the exchange of the perturbation energy between the fluid and the porous layers are derived in Sec.~\ref{sec:governing equations}. The propositions are proved in Sec.~\ref{sec:propositions}. The applications of the propositions to the sliding Couette, annular Poisuille, and Hagen--Poisuille are shown in Sec.~\ref{sec:Applications}. Finally, the major conclusions obtained in the present study are presented in Sec.~\ref{sec:conclusions}.

\section{Governing equations} \label{sec:governing equations}

\begin{figure}
  \centerline{\includegraphics[width=0.65\textwidth]{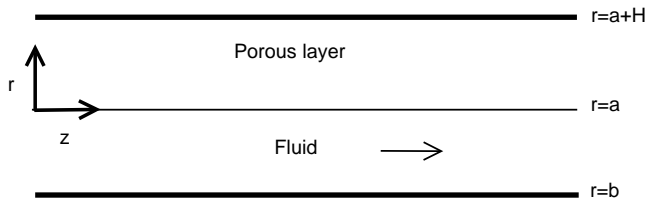}}
  \caption{\small  Side-view schematic of the flow geometry in dimensional coordinates. The fluid flows as a consequence of the driving force imparted by the moving cylinder at $r=b$ or due to an imposed pressure gradient. The fluid-porous layer form an interface at $r=a$ through which the perturbation energy exchange occurs. For Hagen--Poiseuille flow, $b=0$ and $a=R$. Please note that to encompass all possible flows in cylindrical geometries, the figure shows only the side-view of the half flow geometry.  }
  \label{fig:schematic}
\end{figure}

Consider a fluid of constant density $\rho$ and viscosity $\mu$ is flowing through an annular region with the outer wall being a porous layer saturated with the same fluid as shown in figure~\ref{fig:schematic}. Let $U(r)$ and $V(r)$ be the arbitrary axisymmetric base-state velocity profiles  for the fluid and porous layers, respectively.  The fluid layer extends from $r=b$ to $r=a$ while the porous layer extends from $r=a$ to $r=a+H$. Thus, fluid-porous layer form an interface at $r=a$. At $r=a+H$, a rigid impermeable surface is assumed to be present. Let $\mathbf{u}=(u_r,u_\theta,u_z)$ be the averaged velocity field in the porous layer. The continuity equation for the porous layer is
\begin{subequations}
\begin{eqnarray}
\nabla \cdot \mathbf{u} =0,
\label{eq:continuity-equation-porous}
\end{eqnarray}
The fluid flow in the porous layer is described by using the Darcy's model \citep{Nield-1977,Chen-Chen-1988,Chang-et-al-2006,Chang-2005,Chang-2006,Hill-Straughan-2008,Chang-et-al-2017},
\begin{eqnarray}
 \frac{\rho}{\epsilon} \frac{\partial \mathbf{u}}{\partial t} =-\nabla p_m - \frac{\mu}{\kappa} \mathbf{u}, \label{eq:porous-layer-mom}
\end{eqnarray}
where $\epsilon$ and $\kappa$ are the porosity and permeability of the porous layer, respectively and $p_m$ is the interstitial pressure. The viscous term in the above equation is retained since the viscous term for the flow in the porous layer will be considerably higher than that for the fluid at the same external driving force due to the presence of $\kappa \ll 1$ \citep{Childs-Collis-George-1950,Ochoa-Whitaker-1995,Goyeau-et-al-2003,Goharzadeh-et-al-2005} in the denominator.

Additionally, if one were to consider the unsteady Brinkman model for the porous layer then it introduces an additional term $\mu_e \nabla^2 \mathbf{u}$ on the right-hand side of equation~(\ref{eq:porous-layer-mom}) where $\mu_e$ is the effective viscosity given by $\mu_e=\mu/\epsilon$ \citep{Hill-Straughan-2008}. In terms of the length and velocity scales for the porous layer, the term introduced by the Brinkman model is $\mu_e \nabla^2 \mathbf{u}=\mu U_{max}/(\epsilon H^2)$ while the last term on the right-hand side of equation~(\ref{eq:porous-layer-mom}) is $\mu \mathbf{u}/\kappa =\mu U_{max}/\kappa$ where $U_{max}$ is the maximum velocity in the porous layer. A ratio of these two terms is $\mu_e \nabla^2 \mathbf{u}/(\mu \mathbf{u}/\kappa )=\kappa/(\epsilon H^2)=\delta^2/\epsilon \ll 1 $ where $\delta$ is the Darcy number. Therefore, the viscous term introduced by the unsteady Brinkman model will be considerably small compared to the viscous term in the unsteady Darcy equation. A further scaling analysis shows that at a high Reynolds number, the viscous term $\mu \mathbf{u}/\kappa$ will be still comparable to the other terms due to the presence of $\kappa$. However, the Brinkman viscosity term $\mu_e \nabla^2 \mathbf{u}$ can be neglected.  Thus, in the context of the present problem, the unsteady Brinkman equation does not provide an advantage compared to the unsteady Darcy equation.
\label{eq:goveq}
\end{subequations}

\subsection{Perturbation energy exchange term} \label{sec:work exchange term}

To derive the desired results, first, the perturbation energy exchange term between the porous and fluid layers will be derived for which the perturbation energy equation in the porous layer is essential. To obtain the energy equation, the perturbation equation~(\ref{eq:porous-layer-mom}) is multiplied by  $\mathbf{u'}^{*}$ to obtain  
\begin{eqnarray}
 \frac{\rho}{\epsilon} \mathbf{u'}^{*} \cdot \frac{\partial \mathbf{u'}}{\partial t} =- \mathbf{u'}^{*} \cdot \nabla p'_m - \frac{\mu}{\kappa} \mathbf{u'}^{*} \cdot \mathbf{u'}, 
\end{eqnarray}
where superscript $*$ signifies a complex conjugate. The above equation can be further modified using the perturbation form of the porous layer continuity equation~(\ref{eq:continuity-equation-porous}) for the complex conjugate
\begin{eqnarray}
  \frac{\rho}{\epsilon} \mathbf{u'}^{*} \cdot \frac{\partial \mathbf{u'}}{\partial t} =-  \nabla \cdot (p'_m \mathbf{u'}^{*}) - \frac{\mu}{\kappa} \mathbf{u'}^{*} \cdot \mathbf{u'}. 
\end{eqnarray}
Next, the above equation is integrated in the volume ($\nu$) spanned by one wavelength in the $z$-direction, from $0$ to $2\pi$ in the $\theta$-direction and over the thickness of the porous layer in the $r$-direction to obtain
\begin{eqnarray}
  \frac{\rho}{\epsilon} \int_\nu \mathbf{u'}^{*} \cdot \frac{\partial \mathbf{u'}}{\partial t} d\nu =-  \int_s p'_m \mathbf{u'}^{*} \cdot \mathbf{n} ds - \frac{\mu}{\kappa} \int_\nu \mathbf{u'}^{*} \cdot \mathbf{u'} d \nu,
\end{eqnarray}
where $s$ is the area of the surfaces at $r=a$ and $r=a+H$. The divergence theorem has been used for the pressure term to transform the volume integral to the area integral. The quantity $\mathbf{n}$ is the normal to the fluid layer and porous layer interface.

 For the purpose of the linear stability analysis, dynamical quantities such as velocities and pressure are decomposed into the base-state and perturbed state, as $f(\mathbf{x},t)= F(x_3)+f'(\mathbf{x},t)$.
Here, $f(\mathbf{x},t)$  is any dynamic quantity and $F(r)$ and $f'(\mathbf{x},t)$ are the base-state and perturbation parts of the corresponding quantity, respectively. Here, a prime signifies the small perturbation quantity. In the linearised perturbation equations (\ref{eq:goveq}), the normal modes of the following form are then substituted,
\begin{eqnarray}
f'(\mathbf{x},t)=\tilde f(y) e^{i (k z + n \theta - k c t)}, \label{eq:normal-modes}
\end{eqnarray}
where $k$ and $n$ are the (real) wavenumbers and $\tilde f(y)$ is the eigenfunction of $f'(\mathbf{x},t)$. 
The other parameter, $c=c_r+ i c_i$ is the complex phase speed, which characterises the temporal phase speed and growth of the perturbations. 
Therefore, the flow is considered to be temporally unstable if at least one eigenvalue satisfies the condition $ c_i >0$.
 Substituting normal modes~(\ref{eq:normal-modes}) in the above equation leads to
\begin{eqnarray}
 \frac{-ik \rho c}{\epsilon} \int_\nu  \mathbf{ \tilde u}^{*} \cdot  \mathbf{\tilde u} \, d \nu =-  \int_s \tilde p_m \mathbf{ \tilde u}^{*} \cdot \mathbf{n} ds - \frac{\mu}{\kappa} \int_\nu \mathbf{ \tilde u}^{*} \cdot  \mathbf{\tilde u} \, d \nu .
\end{eqnarray}
Let $K_e= \rho/(4\pi \epsilon \lambda)\int_\nu  \mathbf{ \tilde u}^{*} \cdot  \mathbf{\tilde u} \, d \nu$ be the kinetic energy integral and $D_e= \mu/(4\pi \kappa \lambda)\int_\nu  \mathbf{ \tilde u}^{*} \cdot  \mathbf{\tilde u} \, d \nu$ be the viscous dissipation integral for the perturbations in the porous layer where $\lambda = 2 \pi/k$ is the wavelength of the perturbations.
\begin{eqnarray}
 2ik c K_e =    \left[ (\tilde p_m \tilde u^{*}_r)_{r=a} + (\tilde p_m \tilde u^{*}_r)_{r=a+H} \right] + 2 D_e, 
\end{eqnarray}
where the area integral has been integrated. Assuming  an impermeable boundary at $r=a+H$ \citep{Chang-et-al-2006,Hill-Straughan-2008,Chang-et-al-2017}, the perturbations must vanish. Thus, the term $(\tilde p_m \tilde u^{*}_r)_{r=a+H}$ will vanish where subscript indicates the evaluation of the corresponding quantity at the designated $r$ value. Thus, the perturbation energy exchange term between the flowing fluid and the porous layer is,
\begin{eqnarray}
   (\tilde p_m \tilde u^{*}_r)_{r=a} = 2\left(i k c K_e - D_e \right). \label{eq:work term}
\end{eqnarray}

\subsection{Perturbation equation for the fluid}

 Let the velocity field in the fluid be  $\mathbf{v}=(v_r,v_\theta,v_z)$. 
The continuity equation is
\begin{subequations}
\begin{eqnarray}
\nabla \cdot \mathbf{v} =0.
\label{eq:continuity-equation}
\end{eqnarray}
The Euler equation for the fluid is
\begin{eqnarray}
  \rho \frac{\partial \mathbf{v}}{\partial t} + (\mathbf{v}\cdot \nabla) \mathbf{v}  =- \nabla p,
\end{eqnarray}
where $p$ is the pressure field in the fluid. 
\label{eq:goveq fluid}
\end{subequations}
Using normal modes~(\ref{eq:normal-modes}) in (\ref{eq:goveq fluid}) leads to
\begin{subequations}
\begin{eqnarray}
\frac{1}{r} D(r \tilde v_r) + \frac{in}{r} \tilde v_\theta + i k \tilde v_z =0, \label{eq:continuity-eqn-tilde}\\
\rho ik (V-c) \tilde v_r + D \tilde p =0, \label{eq:rmom-tilde}\\
\rho ik (V-c) \tilde v_\theta + \frac{in}{r} \tilde p =0,  \label{eq:thmom-tilde}\\
\rho ik (V-c) \tilde v_z + \rho  DV \tilde v_r + ik \tilde p =0,  \label{eq:zmom-tilde}
\end{eqnarray}
where $D=d/dr$. The perturbation equations for the porous layer after the substitution of the normal modes will not be useful in the further analysis, thus, will not be reported for the sake of brevity. 
\label{eq:tilde-equations-fluid}
\end{subequations}

The above equations~(\ref{eq:tilde-equations-fluid}) for the fluid are further manipulated to yield Rayleigh-like equation for the fluid layer \citep{shankar-kumaran-2000}
\begin{eqnarray}
\frac{d}{dr} \left[ \frac{r}{n^2+k^2 r^2} \frac{d}{dr} (r \tilde v_r) \right] - \tilde v_r - \frac{r \tilde v_r}{V-c} \frac{d}{dr} \left( \frac{r DV}{n^2+k^2 r^2} \right)  = 0. \label{eq:rayleigh-eqn}
\end{eqnarray}
\begin{subequations}
The boundary conditions for the present problem are as follows. At $r=b$, using axisymmetry for the Hagen-Poisuille flow and impermeability for sliding Couette and annular Poiseuille flows yields
\begin{eqnarray}
   \tilde v_r =0.
\end{eqnarray}
At the fluid-porous layer interface ($r=a$), the interface conditions are \citep{Hill-Straughan-2008,Chang-et-al-2017}
\begin{eqnarray}
   \tilde v_r=\tilde u_r,\\
   \tilde p  = \tilde p_m,
\end{eqnarray}
 At $r=a+H$, an impermeable wall is assumed, thus,
\begin{eqnarray}
   \tilde u_r=0.
\end{eqnarray}

\label{eq:bcs-tilde}
\end{subequations}

The perturbation energy exchange term~(\ref{eq:work term}) can be expressed in terms of the fluid layer perturbations by using the normal velocity and pressure continuity conditions~(\ref{eq:bcs-tilde}) at the fluid-porous layer interface to obtain
\begin{eqnarray}
  (\tilde p_m \tilde u^{*}_r)_{r=a} = (\tilde p \tilde v^{*}_r)_{r=a}. 
\end{eqnarray}
To obtain an expression for $\tilde p$, equation~(\ref{eq:thmom-tilde}) is multiplied by $(in/r)$ and  equation~(\ref{eq:zmom-tilde}) is multiplied by $(ik)$. The resulting equations are then added. Using the continuity equation~(\ref{eq:continuity-eqn-tilde}) and substitution $\psi = -ik \tilde v_r/r  $ yields
\begin{eqnarray}
\tilde p = \frac{\rho k^2 r}{n^2+k^2 r^2} \left[ -(V-c) D\psi + DV \psi \right],
\end{eqnarray}
where $\psi$ is the streamfunction. At $r=a$, the perturbation energy exchange term thus becomes
\begin{eqnarray}
(\tilde p_m \tilde u^{*}_r)_{r=1} = \frac{i\rho k^3}{n^2+k^2} \left[ -(V_I-c) D\psi_I \psi^*_I + DV_I |\psi_I|^2 \right], \label{eq:work term fluid quantities}
\end{eqnarray}
where the subscript $I$ signifies the quantities evaluated at fluid-porous layer interface (i.e. at $r=a$). Henceforth, the maximum and minimum velocities of the fluid, $V_{max}$ and  $V_{min}$, respectively are assumed to be positive. \\

 \section{Propositions} \label{sec:propositions}

\subsection{Proposition 1} \label{sec:Proposition 1}

\emph{Statement: An inviscid flow through a porous tube is unstable only if 
\begin{eqnarray}
\frac{d}{dr} \left( \frac{r DV}{n^2+k^2 r^2} \right) (V-V_I)<0,
\end{eqnarray}
somewhere in the flow.}

 \emph{Proof:} The equation~(\ref{eq:rayleigh-eqn}) in terms of the streamfunction $\psi$ is
\begin{eqnarray}
\frac{d}{dr} \left[ \frac{r}{n^2+k^2 r^2} D\psi \right] - \frac{\psi}{r} - \frac{\psi}{V-c} \frac{d}{dr} \left( \frac{r DV}{n^2+k^2 r^2} \right)  = 0. \label{eq:rayleigh-eqn-psi}
\end{eqnarray}
Multiplying the above equation by $\psi^*$ and integrating over the fluid domain leads to
 \begin{eqnarray}
   \int_a^b \left[ \frac{r}{n^2+k^2 r^2} |D\psi|^2+ \frac{|\psi|^2}{r} + \frac{|\psi|^2}{V-c} \frac{d}{dr} \left( \frac{r DV}{n^2+k^2 r^2} \right)   \right] d r = -\frac{\psi^*_I D \psi_I}{n^2+k^2}, \label{eq:rayleigh-eqn-mod-int},
\end{eqnarray}
Using equations~(\ref{eq:work term fluid quantities}) and (\ref{eq:work term}), the above equation becomes
\begin{eqnarray}
\nonumber
   \int_a^b \left[ \frac{r}{n^2+k^2 r^2} |D\psi|^2+ \frac{|\psi|^2}{r} + \frac{|\psi|^2}{V-c} \frac{d}{dr} \left( \frac{r DV}{n^2+k^2 r^2} \right)   \right] d r\\
  = \frac{1}{V_I - c} \left[ \frac{2}{i \rho  k^3} \left(i k c K_e - D_e \right) - \frac{DV_I}{n^2+k^2} |\psi|^2  \right].
\end{eqnarray}
Multiplying above equation by $(V_I - c)$ and taking imaginary part of the resulting equation leads to
\begin{eqnarray}
\nonumber
  c_i    \int_a^b \left[ \frac{r}{n^2+k^2 r^2} |D\psi|^2+ \frac{|\psi|^2}{r} + \frac{|\psi|^2}{|V-c|^2} \frac{d}{dr} \left( \frac{r DV}{n^2+k^2 r^2} \right) (V-V_I)  \right] d r\\
  =  -\frac{2}{\rho k^2} \left( c_i K_e +\frac{D_e}{k}\right) 
\end{eqnarray}
An inspection of the above equation shows that unstable perturbations exist only if 
\begin{eqnarray}
\frac{d}{dr} \left( \frac{r DV}{n^2+k^2 r^2} \right) (V-V_I)<0,
\end{eqnarray}
somewhere in the flow thereby proving the proposition since $K_e$ and $D_e$ are positive-definite. This proposition is equivalent to \cite{Fjortoft-1950} theorem for planar flows past rigid surfaces and \cite{Maslowe-1974,Lalas-1975} condition for the flow in the cylindrical geometries with rigid wall.

\subsection{Proposition 2} \label{sec:Proposition 2}

\emph{Statement: The real part of the phase speed ($c_r$) of the unstable inviscid perturbations must satisfy the condition,}
\begin{eqnarray}
  V_{min} < c_r < V_{max}.
\end{eqnarray}

\emph{Proof:} To obtain the proof, following substitution is made
\begin{eqnarray}
  f=-\frac{\psi}{(V-c)}, \label{eq:f-def}
\end{eqnarray}
upon which the equation~(\ref{eq:rayleigh-eqn}) can be recast as
\begin{eqnarray}
   -\frac{d}{dr} \left[ \frac{r}{n^2+k^2 r^2} (V-c)^2  Df  \right] +  \frac{1}{r} (V-c)^2 f =0.
\end{eqnarray}
Multiplying the above equation by $f^*$ and integrating on $r$ domain,
\begin{eqnarray}
  \int_a^b (V-c)^2 \left[\frac{r}{n^2+k^2 r^2} |Df|^2+ \frac{1}{r} |f|^2 \right] d r = -  \frac{(V_I-c)^2}{n^2+k^2}  f^*_I Df_I. \label{eq:rayleigh-inegrated}
\end{eqnarray}
For the flow past a rigid surface, the term on the right hand side of the above equation will vanish. thus the term $-  \frac{(V_I-c)^2}{n^2+k^2}  f^*_I Df_I$ purely exists due to the presence of the porous layer. Using equation~(\ref{eq:f-def}) and (\ref{eq:work term fluid quantities}) gives
\begin{eqnarray}
  - \frac{(V_I-c)^2}{n^2+k^2}  f^*_I Df_I = \frac{2}{i\rho  k^3 (V_I - c^*)} \left(i k c K_e - D_e \right). 
\end{eqnarray}
Substituting the above equality in equation~(\ref{eq:rayleigh-inegrated}) results in
\begin{eqnarray}
    \int_a^b (V-c)^2 \left[\frac{r}{n^2+k^2 r^2} |Df|^2+ \frac{1}{r} |f|^2 \right] d r = \frac{2}{i\rho  k^3 (V_I - c^*)} \left(i k c K_e - D_e \right). \label{eq:rayleigh-inegrated-2}
\end{eqnarray}
 The imaginary part of the above equation is 
\begin{eqnarray}
\nonumber
  c_i \int_a^b (V-c_r) \left[\frac{r}{n^2+k^2 r^2} |Df|^2+ \frac{1}{r} |f|^2 \right] d r \\
  =\frac{c_r (D_e + 2 k c_i K_e) - (D_e + k c_i K_e) V_I}{ \rho k^3} \frac{1}{|V_I-c|^2}. \label{eq:rayleigh-inegrated-imag}
\end{eqnarray}
 An inspection of the above equation shows that for unstable perturbations $c_i>0$, $c_r > V_I$. However, if $c_r>V_{max}$ then the assumption of $c_i >0$ will contradict due to the left-hand side of the above equation. Thus, for the unstable inviscid modes, $V_{max}$ is the upper bound. A symmetric set of arguments show that the lower bound on $c_r$ for the unstable perturbations is $V_{min}$ thereby proving the proposition. This proposition is equivalent to the bounds derived by Rayleigh for planar flows past rigid surfaces \citep{yeo87}.\\

\subsection{Proposition 3} \label{sec:Proposition 3}

\emph{Statement: The eigenvalues for an inviscid flow past a porous layer satisfy $|c|^2 < max(V^2)$.}

\emph{Proof:} To prove this proposition, multiplying equation~(\ref{eq:rayleigh-inegrated-2}) by $c^*$ and then taking the imaginary part of the resulting equation gives
\begin{eqnarray}
 c_i \int_a^b (V^2-|c|^2) \left[\frac{r}{n^2+k^2 r^2} |Df|^2+ \frac{1}{r} |f|^2 \right] d r = \frac{2 c_i V_I}{\rho k^3 |V_I - c|^2} \left(k c_i K_e + D_e \right) . 
\end{eqnarray}
The quantity on the right-hand side of the above equation is positive for unstable perturbations. For $c_i>0$, the left-hand side also must be positive which implies $|c|^2<V^2$ thereby proving the proposition.\\

\subsection{Proposition 4}. \label{sec:Proposition 4}

\emph{Statement: The temporal growth rate ($k c_i$) of the unstable perturbations satisfies}

\begin{eqnarray}
k c_i \leq \frac{1}{2} max  \left|\frac{k^2 r^2 DV^2}{n^2+k^2 r^2} \right|.
\end{eqnarray}

\emph{Proof:} Similar to Proposition 2 above, to prove this proposition let
\begin{eqnarray}
  g=\frac{\psi}{\sqrt{V-c}}. \label{eq:g-def}
\end{eqnarray}
Upon substitution, the Rayleigh equation~(\ref{eq:rayleigh-eqn}) can be rearranged to give
\begin{eqnarray}
\nonumber
\frac{d}{dr} \left[\frac{r}{n^2+k^2 r^2} (V-c) Dg \right] \\
- g \left[\frac{1}{2} \frac{d}{dr} \left( \frac{r DV}{n^2+k^2 r^2} \right) + \frac{1}{r} (V-c) + \frac{r}{n^2+k^2 r^2} \frac{DV^2}{4(V-c)}      \right] = 0.
\end{eqnarray}
Multiplying the above equation by $g^*$ and integrating over the fluid domain gives
\begin{eqnarray}
\nonumber
 \int_a^b |g|^2 \left[\frac{1}{2} \frac{d}{dr} \left( \frac{r DV}{n^2+k^2 r^2} \right) + \frac{1}{r} (V-c) + \frac{r}{n^2+k^2 r^2} \frac{DV^2}{4(V-c)}      \right] dr \\
 +\int_a^b \left[\frac{r}{n^2+k^2 r^2} (V-c)\right] |Dg|^2 d r
= -\frac{1}{n^2+k^2}(V_I - c) Dg_I g^*_I.
\end{eqnarray}
Using equations~(\ref{eq:work term fluid quantities}) and (\ref{eq:g-def}), the right-hand side of the above equation becomes
\begin{eqnarray}
\nonumber
 \int_a^b |g|^2 \left[\frac{1}{2} \frac{d}{dr} \left( \frac{r DV}{n^2+k^2 r^2} \right) + \frac{1}{r} (V-c) + \frac{r}{n^2+k^2 r^2} \frac{DV^2}{4(V-c)}      \right] dr \\
 \nonumber
 +\int_a^b \left[\frac{r}{n^2+k^2 r^2} (V-c)\right] |Dg|^2 d r\\
= \frac{2}{i\rho  k^3 |V-c|} \left(i k c K_e - D_e \right) - \frac{1}{2(n^2+k^2)} DV_I |g_I|^2.
\end{eqnarray}
Taking the imaginary part of the above equation and multiplying the resulting equation by $k^2 c_i$ gives
\begin{eqnarray}
\nonumber
c_i^2 \int_a^b \frac{k^2 r}{n^2+k^2 r^2} |Dg|^2 d r + \int_a^b |g|^2 \left[  \frac{k^2 c_i^2}{r} - \frac{k^2 r}{n^2+k^2 r^2}\frac{DV^2 c_i^2}{4|V-c|^2} \right] d r \\
= \frac{-2 c_i}{\rho |V-c|} \left( c_i K_e +\frac{D_e}{k}\right).
\end{eqnarray}
 The above equation can be recast for the unstable perturbations $c_i>0$ and noting that $c_i^2 \leq |V-c|^2$ leads to 
\begin{eqnarray}
\nonumber
  \int_a^b \frac{1}{r} |g|^2 \left[  k^2 c_i^2 - \frac{1}{4} max \left|\frac{k^2 r^2 DV^2}{n^2+k^2 r^2} \right| \right] d r \\
\leq \frac{-2 c_i}{\rho |V-c|} \left( c_i K_e +\frac{D_e}{k}\right).
\end{eqnarray}
For unstable perturbations, the right-hand side of the above equation is negative, thus,
\begin{eqnarray}
k c_i \leq \frac{1}{2} max  \left|\frac{k^2 r^2 DV^2}{n^2+k^2 r^2} \right|^{\frac{1}{2}},
\end{eqnarray}
thereby proving the proposition. This proposition is the equivalent of \cite{Hoiland-1953} theorem for the planar flows past rigid surfaces and \cite{Lalas-1975} condition on the growth rate for the flows in cylindrical geometries.\\

\section{Applications} \label{sec:Applications}

In this section, we apply the above-derived propositions to probe inviscid modes of instability in the three types of flows as follows.

\subsection{Sliding Couette flow}

Sliding Couette flow is through an annulus spanning from $r=b$ to $r=a$ and porous wall extending from $r=a+H$. The cylinder at $r=b$ is moving with a steady speed $V_b$. The resultant flow has a logarithmic velocity profile in dimensional form
\begin{eqnarray}
V= A_1  log(r) + A_2,\\
A_1 = \frac{a \, \alpha \, V_b}{\sqrt{\kappa} + a \, \alpha \, log(b/a)},\\
A_2 = \frac{V_b (\sqrt{\kappa} - a \, \alpha \, log(a))}{\sqrt{\kappa} + a \, \alpha \, log(b/a)},
\end{eqnarray}
where $\alpha$ is the \cite{Beavers-Joseph-1967} constant.
Applying Proposition~\ref{sec:Proposition 1},  
\begin{eqnarray}
\frac{d}{dr} \left( \frac{r DV}{n^2+k^2 r^2} \right) (V-V_I) = -\frac{2 \, a \, \alpha \, k^2 \, r \, V_b^2 (-\sqrt{\kappa} + a \, \alpha \, log(a/r))}{(n^2 + k^2 r^2)^2 (-\sqrt{\kappa} + a \, \alpha \, log(a/r))^2}.
\end{eqnarray}
Thus, for the existence of the instability according to Proposition~\ref{sec:Proposition 1} leads to the condition $r<a \, exp\left( -\sqrt{\kappa}/(a \alpha)\right)$. Since fluid domain is in $r\in [b,a]$, thus, the only way for this condition can be satisfied is $\sqrt{\kappa}/(a \, \alpha)>1$ which in dimensionless terms implies $h>A \, \alpha/\delta$. Here, $A=a/(b-a)$, $h=H/(b-a)$, and $\delta=\sqrt{\kappa}/H$ are the dimensionless location of the fluid-porous layer interface, dimensionless thickness of the porous layer, and the Darcy number, respectively. To conclude, there is a minimum thickness of the porous wall required to set in the instability. It must be noted that a planar Couette flow past a Darcy porous layer is linearly stable. However, from the present analysis, the sliding Couette flow past a porous layer is linearly unstable thereby showing the role of the curvature in triggering instability.

\subsection{Hagen--Poiseuille flow} \label{sec:hagen-poi}

The Hagen--Poiseuille flow through a tube with the porous wall has the base-state velocity profile of the form
\begin{subequations}

\begin{eqnarray}
V= \frac{1}{4} A_1  r^2 + A_2,\\
A_1 = \frac{1}{\mu} \frac{dp}{dx},\\
A_2 = \frac{1}{4} A_1 \left[-4 \kappa + \frac{2 \sqrt{\kappa} R}{\alpha} - R^2 \right],
\end{eqnarray}
\label{eq:base-state velocity-hpf}
\end{subequations}
\noindent
where  $dp/dx$ is the imposed pressure gradient. Please note that for Hagen--Poiseuille flow $b=0$ and $a=R$. To apply Proposition~\ref{sec:Proposition 1} the required quantity is  
\begin{eqnarray}
\frac{d}{dr} \left( \frac{r DV}{n^2+k^2 r^2} \right) (V-V_I) = -\frac{4 A_1^2 n^2 r ( R^2 - r^2)}{(n^2 + k^2 r^2)^2}.
\end{eqnarray}
The right-hand side of the above equation is negative or zero in the fluid domain. Thus, by Proposition~\ref{sec:Proposition 1} the flow can exhibit both axisymmetric ($n=0$) and non-axisymmetric ($n\neq 0 $) inviscid modes of instability. The Hagen-Poiseuille flow through a tube with impermeable walls is linearly stable. Thus, the instability predicted here is solely due to the presence of the porous wall.

\subsection{Annular Poiseuille flow}

The annular Poiseuille flow with a porous wall has the base-state velocity profile of the form
\begin{subequations}

\begin{eqnarray}
V= \frac{1}{4} A_1  r^2 + A_2 log(r) + A_3,\\
A_1 = \frac{1}{\mu} \frac{dp}{dx},\\
A_2 = \frac{a A_1}{4}  \frac{(a^2 \alpha - \alpha \, b^2 - 2 \, a \, \sqrt{\kappa} + 4 \, \alpha \, \kappa)}{\sqrt{\kappa} + a \, \alpha \, log(b/a)},\\
A_3 = -\frac{A_1}{4}  \frac{b^2 (\sqrt{\kappa} - a \, \alpha \, log(a)) + a (a^2 \alpha - 2 \, a \, \sqrt{\kappa} + 4 \, \alpha \, \kappa) log(b)}{\sqrt{\kappa} + a \, \alpha \, log(b/a)}.
\end{eqnarray}
\label{eq:base-state velocity-apf}
\end{subequations}
\noindent
The quantity, 
\begin{eqnarray}
\frac{d}{dr} \left( \frac{r DV}{n^2+k^2 r^2} \right) (V-V_I) = -\frac{r (2 \, A_2 k^2 - A_1 n^2) (4\, A_3 + A_1 (4 \kappa + r^2) + 4 A_2 log{r})}{4(n^2 + k^2 r^2)^2}. \,\,
\end{eqnarray}
Unlike the sliding Couette or Hagen-Poiseuille flow, an immediate conclusion about  the stability of annular Poiseuille flow is not straightforward due to the complex structure of the right-hand side of the above equation. However, it can be proved that the annular Poieuille flow is unstable as follows. The stability characteristics of the annular Posieuille flow are bounded by two asymptotic limits. The first asymptotic limit is for $b \rightarrow a$ which implies a plane Poiseuille flow since the cylinders are sufficiently near each other such that the curvature effects can be neglected. But from the analysis of \cite{Chang-et-al-2006,Hill-Straughan-2008,Chang-et-al-2017,Samanta-2017}, the plane Poiseuille flow past a porous layer is unstable. Thus, the annular Poiseuille flow is linearly unstable in the narrow gap limit. The other asymptotic limit is achieved for $b\rightarrow 0$ and finite $a$, i.e., Hagen-Poiseuille flow which from Sec.~\ref{sec:hagen-poi} is linearly unstable. Thus, annular Poiseuille flow with a porous wall is linearly unstable.

\section{Conclusions} \label{sec:conclusions}

The temporal stability of an inviscid flow in cylindrical geometries with a porous wall subjected to non-axisymmetric perturbations is investigated in the present work. Proposition~\ref{sec:Proposition 1} shows that the necessary and sufficient condition for the existence of the instability is $\frac{d}{dr} \left( \frac{r DV}{n^2+k^2 r^2} \right) (V-V_I)<0$ somewhere in the flow thereby deriving the equivalent of \cite{Fjortoft-1950} theorem for planar flows. From Proposition~\ref{sec:Proposition 2}, the real part of the phase speed of the unstable perturbations satisfies $ V_{min} < c_r < V_{max}$. The bound on the absolute value of the eigenvalues is given by $|c|^2 < max(V^2)$ in Proposition~\ref{sec:Proposition 3}. The temporal growth rate of the unstable perturbations is found to satisfy $k c_i \leq \frac{1}{2} max  \left|\frac{k^2 r^2 DV^2}{n^2+k^2 r^2} \right|^{\frac{1}{2}}$ in Proposition~\ref{sec:Proposition 4} which is equivalent of the \cite{Hoiland-1953} theorem.

 The derived propositions are then used to investigate the stability of the sliding Couette, annular Poiseuille, and  Hagen-Poiseuille flows with a porous wall. The analysis reveals the existence of both axisymmetric and non-axisymmetric modes of instability in the sliding Couette flow provided that the dimensionless thickness of the porous wall ($h$) satisfies $h>A\alpha/\delta$. The plane Couette flow past a porous layer described by the unsteady Darcy equation is linearly stable. Thus, the instability predicted in the present study is a result of the curvature present in the sliding Couette flow. The annular Poiseuille flow is also found to be unstable to both the axisymmetric and non-axisymmetric perturbations. The important case, i.e., Hagen-Poiseuille flow is found to be unstable to both the axisymmetric and non-axisymmetric inviscid modes of instability. Thus, the present study, for the first time, shows that all three flows, viz., sliding Couette, annular Poiseuille, and  Hagen-Poiseuille flows with a porous wall are linearly unstable. It must be noted that the Hagen--Poiseuille flow through a rigid tube is linearly stable. Therefore, the instability predicted here is solely due to the porous nature of the tube wall.
 
 To quantify the role of the porous wall in destabilising the flows in cylindrical geometries, a numerical stability analysis needs to be carried out. The numerical stability analysis could also reveal other modes of instability besides the inviscid modes predicted here. Additionally, experiments could also be performed on Hagen--Poiseuille flow through a tube with porous wall to probe the predicted instability. Thus, the numerical stability analysis and experiments to probe the instability predicted in the present study could be the next step in understanding the stability of flows through cylindrical geometries with a porous wall.

\section*{Declaration of interest}

The author reports no conflict of interest.

% \section*{Acknowledgments}

\appendix

\bibliographystyle{jfm}
% Note the spaces between the initials
\bibliography{references}

\end{document}